\documentclass[english,a4paper]{article}
\usepackage{mathptmx}
\usepackage[T1]{fontenc}
\usepackage[latin9]{inputenc}
\usepackage{geometry}
\usepackage{float}
\usepackage{textcomp}
\usepackage{url}
\usepackage{amsthm}
\usepackage{babel}
\usepackage{relsize}
\usepackage{amsmath}

\begin{document}

\title{Four Statements about the Fourth Generation\footnote{CERN-PH-TH/2009-091}}

\author{B. Holdom,$^{a}$ W.S. Hou,$^{b}$ T. Hurth,$^{c}$ M. Mangano,$^{c}$
S. Sultansoy,$^{d}$ G. \"Unel$^{e}$}

\date{\emph{Summary of the ``Beyond the 3-generation SM in the LHC era''
 Workshop,\\
CERN, September 4-5, 2008}}

\maketitle
\noindent
\emph{$^{a}$}\emph{\small University of Toronto, Canada} \\
\emph{$^{b}$}\emph{\small National Taiwan University, Taiwan}\\
\emph{$^{c}$}\emph{\small CERN, Switzerland} \\
\emph{$^{d}$}\emph{\small TOBB University of Economics \& Technology, Turkey}\\
\emph{$^{e}$}\emph{\small University of California at Irvine, USA}

\begin{abstract}
This summary of the Workshop ``Beyond the 3-generation SM in the LHC era''
presents a brief discussion of the following four statements about the
fourth generation: 1) It is not excluded by EW precision data;
2) It addresses some of the currently open questions; 3) It can
accommodate emerging possible hints of new physics; 4) LHC has the
potential to discover or fully exclude it.

\end{abstract}

\section*{Introduction} 

It is now generally accepted that the Standard Model (SM) consists of
three fermion families, or generations. However the number of
generations is not fixed by the theory. The asymptotic freedom
constraint from QCD only limits the number of generations to be less
than 9. Neutrino counting at the $Z$ pole shows that the number of
generations with light neutrinos ($m_{\nu}\ll m_{Z}/2$ ) is equal to
3, but neutrino oscillations suggest a new mass scale that is beyond
the SM, and the possibility of additional heavier neutrinos cannot be
excluded. In the era of the LHC, the possibility of the SM with a
fourth generation (SM4) should therefore not be forgotten (for earlier
reviews see~\cite{review}, 
and for more recent work see for example \cite{recent}).

By SM4, we mean a sequential repetition of the existing generation
pattern to 4 quark and 4 lepton left-handed doublets
and corresponding right handed singlets. We use the
commonly known primed notation, i.e. $t'$ and $b'$ for fourth
generation  quarks, and $\tau'$ and $\nu'_\tau$ for the heavy
charged and neutral leptons. The current 95\% CL mass limits from the
PDG are \cite{PDG},
\begin{eqnarray}
m_{t'} > 256\;\hbox{GeV};
 && m_{b'} > 128\;\hbox{GeV}\
    (\hbox{CC decay; 199\;GeV}\;\hbox{for\;100\%\;NC decay});
 \label{qbound} \\
m_{\tau'} > 100.8\;\hbox{GeV};
 && m_{\nu'_\tau} > 90.3\;\hbox{GeV}\
    (\hbox{Dirac\;coupling; 80.5\;GeV\;for\;Majorana\;coupling})
 \label{lbound}
\end{eqnarray}

The following text is a summary of the thematic Workshop ``\emph{Beyond the
3-generation SM in the LHC era''}, on the physics of the SM
with $N>3$ fermion generations, held at CERN on
4-5 September~\cite{wshop}. Besides reviewing the theory, as
well as flavour factory, collider and astroparticle/cosmology
aspects, the aim was to stimulate discussions by bringing together
theorists and experimentalists working on, or interested in, the
subject. The imminent LHC start up placed an emphasis on collider
and flavour physics, especially on the preparation for LHC data
exploitation.

\section*{Statement 1: The fourth generation is not excluded by EW precision data.}

\subsection*{EW precision data} 

The ``oblique parameters'' $S$, $T$ and $U$ provide stringent
constraints on SM4 \cite{STU}. The PDG states that ``An extra generation of
ordinary fermions is excluded at the 6$\sigma$ level on the basis
of the $S$ parameter alone'' \cite{PDG}. The caution that is often not noted
or remembered is that ``This result assumes that ... any new
families are degenerate'', while ``the restriction can be relaxed by
allowing $T$ to vary as well''. \footnote{The issue has also been reopened by a
recent study \cite{kribs}.} The
contribution of the 4th generation fermions (much heavier than the
$Z$ boson) to $S$ is
\begin{eqnarray}
\delta S
 & = & \frac{2}{3\pi}
     - \frac{1}{3\pi}\left[\log\frac{m_{t'}}{m_{b'}}
                         - \log\frac{m_{\nu'_\tau}}{m_{\tau'}}\right],
\label{S}
\end{eqnarray}
where the leading term causes trouble with EW precision data for
degenerate fermion masses.  However, if $m_{t'} > m_{b'}$ and/or
$m_{\nu'_\tau} < m_{\tau'}$ then the $S$ parameter constraint can be
softened. In fact this points towards a heavier Higgs. As the Higgs
mass is raised, a positive contribution to $T$ is required from some
other source.  Mass splitting between heavy fermion doublets provides
such a source and this, from (\ref{S}), helps to alleviate the problem
with $S$.\footnote{After completion of this document, a
  paper by M. S.  Chanowitz~\cite{Chanowitz} appeared, taking CKM
  mixing into account in electroweak precision constraints. It was
  found there that a mixing between 3$^{rd}$ and 4$^{th}$ generations of
  the size of the Cabibbo mixing is allowed.}

When the mass of the heavy neutral lepton $\nu'_\tau$ is close to
$m_{Z}/2$, threshold effects to the $Z$ boson self energy need to
be incorporated. In such an analysis, it is found that the quality of
fit for one extra generation can be the same as the 3 generation
case for certain mass values, while the upper bound on Higgs mass
from the SM fit is largely removed \cite{vysotsky}.

The above discussion assumes that $\nu'_\tau$ has a Dirac mass.
But $\nu'_\tau$ could instead have a Majorana mass along with the
three much lighter neutrinos (and then all four right-handed
neutrinos could have masses well above the electroweak scale). A
dynamical Majorana mass for $\nu'_{\tau L}$ produces an additional
finite and negative contribution to $T$~\cite{stu-majo}. Thus
further splitting in the heavy fermion doublets is needed to
produce a compensating positive contribution to $T$. This then has
the same effect as increasing the Higgs mass in alleviating the
problem with $S$.

\subsection*{CKM unitarity } 

Current measurement errors of the CKM quark mixing (and similarly
PMNS for lepton sector) matrix elements leave ample room for the
possible extension from $3\times 3$ to $4\times 4$.
For example, the most precisely measured first row gives $\vert
V_{ud}\vert^2 + \vert V_{us}\vert^2 + \vert V_{ub}\vert^2 = 0.9999
\pm 0.0011$ \cite{PDG}. If this is equated to $1 - \vert V_{ub'}\vert^2$, one
finds (one-sided 95\% CL),
\begin{eqnarray}
\vert V_{ub'}\vert & < & 0.04. 
 \label{CKMunitarity}
\end{eqnarray}
Note that the bound is much larger than $\vert V_{ub}\vert \sim
0.004$ (the value of which did not matter in the above analysis).
For the second row, $\vert V_{cs}\vert$ is far from well measured.
Improvement is made by using $W$ boson leptonic branching ratio
(where 3 generation lepton unitarity is typically assumed, which
is another cautionary note), giving $\vert V_{cd}\vert^2 + \vert
V_{cs}\vert^2 + \vert V_{cb}\vert^2 = 1.002 \pm 0.027$. The large
errors again tolerate a $\vert V_{cb'}\vert$ value considerably
larger than $\vert V_{cb}\vert \simeq 0.04$. A more rigorous
analysis could be performed by studying the $\tau$ and $W$ decays,
which would yield the correlation between CKM and PMNS
matrices~\cite{HL}. Additionally, an increase of available data
would improve the constraints.

\section*{Statement 2: SM4 addresses some of the currently open questions.}
\subsection*{New CPV source for BAU problem } 

The B factories have confirmed the source of CP violation (CPV) in
the 3 generation KM model. However, it is commonly said that the
KM model (hence SM) falls short from what is needed to satisfy the
Sakharov conditions for generating the baryon asymmetry of the
Universe (BAU), by more than 10 orders of magnitude. This can be
seen from the Jarlskog rephasing invariant measure of
CPV~\cite{Jarlskog},
\begin{eqnarray}
J &=& (m_t^2 - m_u^2)(m_t^2 -  m_c^2)(m_c^2 - m_u^2)
      (m_b^2 - m_d^2)(m_b^2 -  m_s^2)(m_s^2 - m_d^2)\, A,
 \nonumber 
\end{eqnarray}
where $A$ is twice the area of any triangle formed from the
$3\times 3$ SM unitarity condition $V^\dag V = I$. The CPV
triangle area $A \simeq 3\times 10^{-5}$ is small, but the
stronger suppression factors are the small masses, or Yukawa
couplings, of SM quarks other than the top, as compared to the
v.e.v. scale.

From the last point, however, it was recently pointed
out~\cite{bau} that, in SM4, most of the mass suppressions can be
bypassed in the (generalized) Jarlskog invariant, while bringing
in a larger CPV ``triangle'' area that could impact on $b\to s$
transitions (see below). The cumulative gain is of order $10^{13}$
to $10^{15}$.

While this is not a proof, and issues such as the order of the phase
transition remain, it is exciting to note that the KM structure
with 4 generations may provide enough CPV for the matter-dominated
Universe.

\subsection*{New perspectives on the Higgs naturalness problem}

The heavy quark loops will contribute new terms in the running of
the quartic coupling that are proportional to powers of the new
Yukawa couplings, namely $\mu d\lambda/d\mu\propto \lambda\,
y_{q'}^2-y_{q'}^4+...$ . The result is to produce a smaller
allowed range for the Higgs mass (increasing lower limit and
decreasing upper limit) if one wants the quartic coupling to
remain finite and positive at 1 TeV. This allowed range decreases
quite quickly as the fourth generation masses are increased. Even more
dramatic is the large contribution to the Higgs mass from the
heavy fermion loops, $\delta m_h^2\approx \left[m_{q'}/\mbox{400
GeV}\right]^2\Lambda^2$. Here $\Lambda$ represents the scale of
new physics needed to cut off these loops, which is at least as
large as $m_{q'}$. Finally, there are the large Yukawa couplings
themselves that run quickly and lead to Landau poles not far above
a TeV. While supersymmetry could be invoked to control the heavy
fermion loop effects, the running of the Yukawa couplings is not
so easy to control.
Given these considerations, we see that the discovery
of a 4th generation would impact directly on the likelihood of
the simplest realization of electroweak symmetry breaking,
namely the Higgs boson.  This is especially true if the 4th
generation quarks have masses as high as $\sim 600$ GeV, the
so-called unitarity bound. The Goldstone bosons of electroweak
symmetry breaking would then couple so strongly to these quarks
that the concept of an elementary scalar field is no longer
appropriate.

Instead, one would conclude that some strong
interaction is responsible for electroweak symmetry breaking via a
condensation of the fourth generation fermions.
What this dynamics is and how it links to the light fermions
remains to be determined. For instance, dynamically broken gauge
interactions could not only be responsible for the condensation,
but could also connect different generations and feed mass from heavy
to light fermions \cite{dyn_sym}. In this case the new flavour physics should
range from a TeV up to about 1000 TeV, and the new physics has
large (small) impact on heavy (light) fermions. Sufficiently light
masses for three neutrinos can follow if right handed neutrinos
masses are $\sim1000$ TeV. A light remnant of these new flavour
interactions, the X boson, should also couple to the third generation,
and can be searched for at the LHC via, e.g. $b\bar{b}\to
X\to\tau^{+}\tau^{-}$ \cite{x-boson}.

Alternatively, in a model with fermions propagating in a
five-dimensional AdS space, it is possible to break the electroweak
symmetry via the condensation of the fourth generation, driven by
their interactions with the Kaluza-Klein gauge bosons and by the
presence of bulk higher-dimensional operators \cite{5D_AdS}.
This dynamical mechanism results in a heavy composite Higgs which
is highly localized towards the infrared boundary (brane). The
localization of the fermions in the 5D bulk gives rise to the Yukawa
couplings. This picture is complementary, and may provide insight to,
the 4D view of strong interactions.

\subsection*{New perspectives into the fermion mass hierarchy problem}

In the Standard Model, the Yukawa couplings spread over an
unnaturally wide range of values, exhibiting a similar hierarchy
for different type of fermion charges. It may be more natural to
assume that all the fermion-Higgs couplings are of the same order,
yielding a single non-zero eigenvalue, $M_{44}$, of the fermionic
mass matrix $M$. This idea is known as the flavour democracy or
Democratic Mass Matrix (DMM) approach \cite{DMM}. In such a  scenario,
the observed masses of
fermions in the first three generations arise from perturbations to
$M$. The mass differences among the third generation charged
fermions do not allow such a parameterization for a $3\times3$
mass matrix. Therefore, flavour democracy requires a 4th SM
generation~\cite{FLDreview}. This would
 also favor smaller masses for the known neutrinos~\cite{Silva}.

\subsection*{New inroads into the Dark Matter problem } 

In a number of models, like the Pati-Salam model or in simpler
phenomenological approaches, there are additional fermions such as
mirror fermions, singlets and sterile neutrinos, even stable
quarks \cite{Hung}. The new fermions could provide answers to a number of
astrophysics problems such as Dark Matter (DM), pulsar kicks etc.
There are group theoretical arguments based on spin-charge
unification or spin-generation considerations that predict the number
of SM like generations as 8, with the additional fermions as DM
candidates \cite{mankoc}. These additional fermions could either decouple
completely from their SM counterparts and serve as cold DM, or
they may be stable particles that form bound neutral atomic states
and serve as composite DM.

For example, some heavy charge $+2/3$ quarks,
denoted as $U$ to emphasize their stability, could
remain in the early Universe after BBN.
Such quarks could form $\overline{U}\overline{U}\overline{U}$
hadrons, which could bind with primordial helium into atom-like
"O-helium" states. In this scenario, O-helium atoms decouple from
plasma and radiation before recombination, and play the role of
warm DM \cite{Khlopov}. Although the interaction of O-helium with terrestrial
matter would slow it down to below the underground DM direct
search threshold, it can be searched for in ground-based
experiments and in space. Annual modulations of ionization signal
from $\overline{U}\overline{U}\overline{U}$ captured by
$^{53}I_{127}$ and $^{82}Tl_{205}$ can explain the results of
DAMA/NaI and DAMA/Libra experiments \cite{dama}, while positrons from
de-excitation of O-helium, excited in its mutual collisions in the
galactic bulge, could explain the excessive positron
annihilation line observed by the Integral experiment.

Searches for new fermions at the upcoming accelerators could check
the validity of these models, hence probe DM.

\section*{Statement 3: SM4 can accommodate emerging possible hints of new physics.}

\subsection*{Tevatron direct search} 

Fourth generation quarks continue to be searched for at the
Tevatron by the CDF and D$\emptyset$ collaborations, as they
continue to collect data. CDF has searched for $t'$ in the charged
current decay $t'\to qW$, i.e. without invoking $b$-tagging \cite{CDFt}. The
$p\bar p \to qqWW$ event gives rise to the signature of $\ell+$
jets $+$ MET (missing transverse mass). The observed slight excess of
events at high $m_{reco}$ (jargon for ``reconstructed'' mass) values
could hint at a new quark heavier than the top, although such an
interpretation would require a production cross section larger than
expected from purely SM couplings of the $t'$.
Both
experiments have also searched for new quarks in the neutral
current decay channels \cite{Demina}. The searches yielded null results
consistent with the non-existence of the FCNCs in the SM.

One should note that the experimental searches usually assume
100\% branching fraction of the mode under study, and tacitly that
the heavy quark is unstable. The exclusion is clearly softened
when the branching fraction or lifetime is affected by mixing
angles, and e.g. the $b'$ and $t'$ mass difference. In principle,
one heavy quark could even become stable in certain special cases,
such as a mixing angle of the order of 10$^{-8}$. Such small
angles could be motivated by a broken discrete symmetry between
generations that only gets restored at the Planck scale. When
quoting the experimental results, the assumptions made have to be
kept in mind~\cite{Sher}.

\subsection*{CPV in $B_{s}\to J/\psi\,\phi$ at Tevatron}

A potentially exciting development emerged at the Tevatron during
2008: there is now a hint for mixing-dependent CPV in $B_{s}\to
J/\psi\phi$~\cite{Fernandez}. CDF and D$\emptyset$ have altogether
conducted three measurements so far, all of which indicate that
$\sin2\Phi_{B_s} = -\sin2\beta_s$ is large and negative, with
central value around $-0.6$ and a significance of roughly
2.8$\sigma$. The SM expectation is $-0.04$. If the central value
stays, it seems that the Tevatron could establish the effect with
the 2010 dataset.

In SM4, with strong $m_{t'}$ dependence (called nondecoupling) in
the box diagram just like the top, and with the new CKM product
$V_{t's}^*V_{t'b}$ bringing in a new CPV phase, the 4th generation
offers the simplest explanation for large deviations of
$\sin2\Phi_{B_s}$ from the SM prediction. In fact, two
predictions~\cite{HNS05,HNS} were made beforehand for
$\sin2\Phi_{B_s} < 0$. The stronger prediction~\cite{HNS} of
$-0.5$ to $-0.7$ was put forth with the CDF observation of $B_s$
mixing in 2006. The argument was that, because typical
$f_{B_s}\sqrt{B_{B_s}}$ values give rise to $\Delta m_{B_s}$
values that are larger than observed, together with the
nondecoupling effect of the $t'$ quark, large and negative
$\sin2\Phi_{B_s}$ would generally follow.

\subsection*{Hints from B-factories} 

The difference in direct CPV (DCPV) measured~\cite{Chang} in
$B^+\to K^+\pi^0$ and $B^0\to K^+\pi^-$ decays by the B-factories,
is now established beyond $5\sigma$ level. It is larger than the
$-10\%$ found in the latter mode. Since the two processes differ
only in the ``spectator'' quark, this dramatic difference was not
anticipated. "In the mind of many~\cite{Peskin,Marco,Li}, however, a
large enhancement of the color-suppressed amplitude could reduce  the
usefulness of this mode as a probe of New Physics". 

It was pointed out~\cite{HNS05,Soni}, nevertheless, that if the
electroweak or $Z$ penguin is any fraction of a culprit, then New
Physics is necessary in the $bsZ$ penguin loop (the $Z$ can
produce a $\pi^0$ but not the $\pi^-$). From the insight that the
top quark is nondecoupled in this loop, just like in the $B_s$
mixing box diagram, introducing the $t'$ quark brings in new CPV
phase via $V_{t's}^*V_{t'b}$. The link to $B_s$ mixing led to the
first prediction, in 2005, that $\sin2\Phi_{B_s} = -\sin2\beta_s$
would be large and negative, a prediction that is in better
agreement with recent CDF and D$\emptyset$ measurements discussed
earlier, compared to 3-generation SM expectations.


\section*{Statement 4: LHC has the potential to discover or fully exclude SM4. }

\subsection*{ATLAS and CMS discovery prospects}

Since the heavy quark pair production cross section at the LHC is
rather large, the ATLAS and CMS experiments have the potential to
discover the 4th generation quarks if they exist. Since partial
wave unitarity gives an upper bound of about 1 TeV to the 4th
generation fermion masses~\cite{partial-wave-unitarity}, a
non-discovery could also mean full exclusion of the SM4 model,
assuming usual decays and mixings \cite{Sher}.

The prominent decay channels depend on the 4$\times$4 CKM matrix,
as well as the $b'$ and $t'$ mass difference. The case with
dominant mixing between the 3rd and 4th generations has been
investigated in the ATLAS TDR, and more recently by the CMS
experiment. The decay channels studied are $t'\to bW$ and $b'\to
tW^-\to bW^+W^-$. With 100 fb$^{-1}$ data, ATLAS claimed 61 (13.5)
$\sigma$ discovery by reconstructing the hadronic decay of the
$t'$ quark pairs of mass of 320 (640) GeV \cite{atlas-tdr}. The study found the
fully hadronic mode of $t'$ and the reconstruction of $b'$ to be
rather difficult.
A recent study by CMS aimed for early physics, searching for same
sign dilepton or trilepton signals from $b'\bar b' \to b\bar b
W^+W^-W^+W^-$.
Using the HT variable (HT$\equiv\sum p_{T}^{i}$) and incorporating
systematical errors, the study found 7.5 (2.0) $\sigma$
significance for a $b'$ quark of mass 300 (400) GeV, with just 100
pb$^{-1}$ at 14 TeV \cite{Chao}. The significance would of course weaken for
10 TeV. Note that the signature of same sign dileptons, as well as the
$t\bar tWW$ final state, is similar to the heavy ``top partner"
quarks considered in~\cite{Contino}.


If the fourth generation quarks prefer to mix with the first two
generation members, the decays $Q\to qW$ where $Q=t',\;b'$ and $q$ is
a light quark (jet), should be considered. Note that the DMM
approach implies nearly equal $t'$ and $b'$ masses, which is
supported by the precision data. If $t'$ and $b'$ are within 50
GeV of each other, they could be hard to distinguish, and the
signal is doubled. Such $t'$ and $b'$ quarks with mass $\sim 500$
GeV can be discovered at 5$\sigma$ significance with 400 pb$^{-1}$
data~\cite{Ozcan}.

Similar to the boosted top, for $W$ bosons with $p_{T} \sim 250$
GeV or higher, one should consider the new tool of reconstructing
the $W$ boson invariant mass as a ``single jet''~\cite{singlejet}.

\subsection*{Other LHC searches} 

The resonant production of $t'$ and $b'$ quarks are studied via the
anomalous processes $gq^{i}\to t'$ and $gq^{j}\to b'$ (where
$q^{i}=u,c$ and $q^{j}=d,s,b$) at the LHC. Such processes are rather
suppressed in SM, but could be induced by the large mass of the fourth
generation quarks. With 10 fb$^{-1}$ of integrated luminosity, the
sensitivity to anomalous couplings, namely the minimum value of
$\kappa/\Lambda$ that can be probed, is 0.01 TeV$^{-1}$~\cite{cakir}.

The aforementioned stable $U$ quark, or other exotic possibilities
for stable heavy quarks, could form heavy hadrons. Then the detection
prospect could be enhanced or suppressed by their interactions
with detector material~\cite{Milstead}.

\subsection*{Impact on Higgs searches at LHC and Tevatron}

The dominant production mechanism of the Higgs boson at hadron
colliders, gluon-gluon fusion, probes heavy quarks in a triangular
loop. The 4th generation quarks lead to an enhancement of this
process.  However, to make a full analysis one should also
recalculate the Higgs branching ratios. Above the $t'\bar t'$ or
$b'\bar b'$ threshold, the 4th generation final state would
dominate over the $t\bar{t}$, $WW$, $ZZ$. If below threshold,
while the same enhancement persists, it is possible to have a
Higgs boson decaying dominantly to ``invisible'' 4th generation
neutral leptons~\cite{belotsky}. The branching ratio for the
$H\rightarrow\gamma\gamma$ channel, however, is in general
reduced, since the extra fourth generation quarks in the
loop tend to cancel the vector boson contribution.

It is possible to even discover $\nu'_\tau$ through its coupling
to the $Z$ and (heavy) Higgs bosons, namely via $pp \to Z/h \to
\nu'_\tau\bar\nu'_\tau \to \mu W\mu W$. This analysis has also the
potential to reveal the Dirac or Majorana nature of $\nu'_\tau$,
with the Majorana case more promising with few fb$^{-1}$
luminosity~\cite{cuhadar}.

The LHC experiments can discover the Higgs via the "golden mode"
($gg\to h \to ZZ \to 4 \ell$) for most of the mass range with
$\sim$ 1~fb$^{-1}$, because of the enhanced cross section due to
SM4~\cite{cetin}. The measured enhancement would provide indirect
evidence about the existence of the fourth generation.
This effect also increases the importance of the $gg\to H\to
WW\to\ell\nu\ell\nu$ search channel at the Tevatron. The
enhancement due to fourth generation quarks is about a factor of 8
for $100<m_{H}<200$ GeV. Although the latest data has not yet been
combined, the Higgs boson is excluded at 95\% level in the range
130 to 190 GeV by CDF alone if the 4th generation exists~\cite{haas}.

\subsection*{LHCb prospects} 

With tantalizing hints for deviation from SM in the time-dependent CPV
measurement of $B_s\to J/\psi \phi$ at the Tevatron, the LHCb
experiment is awaiting data to confirm the effect or reject it.
Even if the SM value of $\sin2\Phi_{B_s} = -\sin2\beta_s \simeq
-0.04$ is correct, LHCb is expected to measure it with just
0.5~fb$^{-1}$ data~\cite{Vagnoni}, and in the process, explore the
full range of New Physics possibilities. Another approach would be
to measure the closure of the unitarity triangle, namely the
measurement of $\gamma/\phi_3$ in the SM. Again, LHCb can play here a
conclusive role.

There is another  slight hint of a deviation from the SM at the B factories,
namely the forward-backward asymmetries
in $B\to K^*\ell^+\ell^-$~\cite{Chang,Babarafb,Belleafb}. 
With the B factories drawing to a close, here
again LHCb could make dramatic impact. With  about 1~fb$^{-1}$
data, LHCb could measure~\cite{Serra} the SM expectation for
$A_{\rm FB}$ in $B_d\to K^*\mu^+\mu^-$ (in particular, check the
zero), or confirm that one again has a deviation. Already with $2-3 fb^{-1}$
a full angular analysis of this mode will
allow for the measurement of the recently proposed new
observables~\cite{LHCbangular} with even
higher NP sensitivity.

It should be clear that the indirect path of probing the fourth
generation through virtual effects cannot provide  all information.
Once again, the direct search prowess of the LHC should be highlighted.

\subsection*{Prospects for future colliders } 

Future linear $e^+e^-$ colliders are especially important for
understanding the leptonic sector of the fourth generation, and for
making precision measurements \cite{Richard}. If kinematics
allows, fourth generation fermions would be pair-produced. For
example, an LC operating at 500 GeV would give very powerful
complements for finding the fourth generation leptons, and for
investigating the light Higgs case \cite{SultansoyFC}. If
sufficiently high $E_{\rm CM}$ cannot be made available, one may
have to search for single production, such as $e^+e^- \to
\tau'\bar\tau$~\cite{SherFC}. But the cross section is not
promising, unless one brings in further New Physics such as 2HDM.



\section*{Acknowledgments}
We thank all participants for their contributions, and for the lively
atmosphere and discussions that their presentations stimulated.
We also thank Dr. Livio Mapelli for his
encouragements and for his participation to the organization of the
workshop. The work of BH is supported in part by the Natural Sciences and
Engineering Research Council of Canada. 
The work of WSH is supported in part be the National Science 
	Council as well as the National Center for Theoretical 
	Sciences (North) of Taiwan.
The work of TH and MLM is supported in part by the European Community's 
Marie-Curie Research Training Network HEPTOOLS under contract 
MRTN-CT-2006-035505. The work of SS is supported by the Turkish Atomic
Energy Authority under the grant number DPT05K120010. GU's work is
supported by the U.S. Department of Energy Grant DE FG0291ER40679 and
by the National Science Foundation Grant PHY-06-12811.


\begin{thebibliography}{10}



 


\bibitem{review}
P.~H.~Frampton, P.~Q.~Hung and M.~Sher,
Phys.\ Rept.\  {\bf 330} (2000) 263. \\
S. Sultansoy, Turk. J. Phys. 22 (1998) 575 [AU-HEP-97-04, Jun 1997].

\bibitem{recent}
A.~Arhrib and W.~S.~Hou,
Phys.\ Rev.\  D {\bf 64} (2001) 073016.\\
D.~Choudhury, T.~M.~P.~Tait and C.~E.~M.~Wagner,
Phys.\ Rev.\  D {\bf 65} (2002) 053002.\\
J.~E.~Dubicki and C.~D.~Froggatt,
Phys.\ Lett.\  B {\bf 567} (2003) 46.\\
L.~Solmaz,
Phys.\ Rev.\  D {\bf 69} (2004) 015003.\\
A.~T.~Alan, A.~Senol and O.~Cakir,
Europhys.\ Lett.\  {\bf 66} (2004) 657.\\
A.~Arhrib and W.~S.~Hou,
JHEP {\bf 0607} (2006) 009.\\
J.~Alwall {\it et al.},
Eur.\ Phys.\ J.\  C {\bf 49} (2007) 791.\\
R.~Fok and G.~D.~Kribs,
Phys.\ Rev.\  D {\bf 78} (2008) 075023.\\
A.~K.~Alok, A.~Dighe and S.~Ray,
arXiv:0811.1186 [hep-ph].\\
Y.~Kikukawa, M.~Kohda and J.~Yasuda,
arXiv:0901.1962 [hep-ph].\\
M.~Bobrowski, A.~Lenz, J.~Riedl and J.~Rohrwild,
arXiv:0902.4883 [hep-ph].\\
O.~Antipin, M.~Heikinheimo and K.~Tuominen,
arXiv:0905.0622 [hep-ph].\\
M.~T.~Frandsen, I.~Masina and F.~Sannino,
arXiv:0905.1331 [hep-ph].




\bibitem{PDG} 
C.~Amsler {\it et al.}  [Particle Data Group],
  Phys.\ Lett.\  B667 (2008) 1.
\bibitem{wshop}
See {\tt
http://indico.cern.ch/conferenceDisplay.py?confId=33285} for the
full agenda and for the talks referred to in the text.
\bibitem{STU} D.~C.~Kennedy and B.W.~Lynn, Nuc. Phys. B322 (1989) 1;
E.~Gates and J.~Terning, Phys. Rev. Lett. 67 (1991) 1840;
T.~Appelquist and J.~Terning, Phys. Lett. B315 (1993) 139. For a
complete review, see M.~E.~Peskin and T.~Takeuchi, Phys. Rev. D46
(1992) 381.
\bibitem{kribs} G.~D.~Kribs, T.~Plehn, M.~Spannowsky, and T.~M.~P.~Tait,
Phys.\ Rev.\ D76 (2007) 075016.
\bibitem{Chanowitz}
M.~S.~Chanowitz,
  arXiv:0904.3570 [hep-ph].
\bibitem{vysotsky} M.~Maltoni, V.~A.~Novikov, L.~B.~Okun, A.~N.~Rozanov, and
M.~I.~Vysotsky, Phys. Lett. B476 (2000) 107; and talk by M. Vysotsky
at this workshop, as documented in 
  V.~A.~Novikov, A.~N.~Rozanov and M.~I.~Vysotsky,
  arXiv:0904.4570 [hep-ph].
\bibitem{stu-majo} B.~Holdom, Phys. Rev. D54 (1996) 721; and talk at this workshop;
for the case of a Dirac mass plus a Majorana mass for
$\nu'_{\tau R}$ see B.~A.~Kniehl and H.~G.~Kohrs, Phys. Rev. D48 (1993) 225.


\bibitem{HL} See talks by E.~Ozcan and H.~Lacker at this workshop, and the references therein.

\bibitem{Jarlskog} C.~Jarlskog, Phys.\ Rev.\ Lett.\  55 (1985) 1039.
\bibitem{bau} W.~S.~Hou, Chin. J. Phys. 47 (2009) 134; and talk at this workshop.
\bibitem{dyn_sym} B.~Holdom, JHEP 0608 (2006) 076.
\bibitem{x-boson} B.~Holdom, Phys. Lett. B666 (2008) 77; and talk at this workshop.
\bibitem{5D_AdS} G.~Burdman, L.~Da~Rold, JHEP 0712 (2007) 086; and talk by L.~Da~Rold at this workshop.

\bibitem{DMM} S.~Sultansoy, AIP conf. proc. {899} (2007)
49 {[}arXiv:hep-ph/0610279{]} and references therein; and talk at this workshop.
\bibitem{FLDreview} H.~Fritzsch, Phys. Lett. B289 (1992) 92;
A.~Datta, Pramana 40 (1993) L503; A.~Celikel, A.~Ciftci,
S.~Sultansoy, Phys. Lett. B342 (1995) 257.

\bibitem{Silva}
J.~I.~ Silva-Marcos, JHEP 12 (2002) 036.
\bibitem{Hung}  P. Q. Hung, Phys. Lett. B649 (2007) 275; and talk at this workshop.
\bibitem{mankoc} A. B. Bracic and N. S. M. Borstnik, Phys. Rev. D74 (2006) 073013;
and talk by N. Borstnik at this workshop.
\bibitem{Khlopov} K. M. Belotsky {\it et al.}, Grav. Cosmol. 11 (2005) 3; and talk by M. Khlopov at this workshop.
\bibitem{dama} R. Bernabei {\it et al.} [DAMA Collaboration],
Nucl. Instrum. Meth. A592 (2008) 297.
\bibitem{CDFt} CDF Collaboration, CDF Note 9446 (2008); and talk by A. Lister at this workshop.
\bibitem{Demina} Talk by R. Demina at this workshop.
\bibitem{Sher} P. Q. Hung, M. Sher,  Phys.\ Rev.\  D77 (2008) 037302; and talk by M. Sher at this workshop.

\bibitem{Fernandez} T. Aaltonen {\it et al.} [CDF Collaboration], Phys. Rev. Lett. 100
(2008) 161802; V. M. Abazov {\it et al.} [D$\emptyset$
Collaboration], Phys. Rev. Lett. 101 (2008) 241801; and talk by J.
P. Fernandez at this workshop.
\bibitem{HNS05} W. S. Hou, M. Nagashima, and A. Soddu,
Phys. Rev. Lett. 95 (2005) 141601; Phys. Rev. D72 (2005) 115007.
\bibitem{HNS} W. S. Hou, M. Nagashima, and A. Soddu,
Phys. Rev. D 76 (2007) 016004; and talk by W. S. Hou at this
workshop.
\bibitem{Chang} S. W. Lin, Y. Unno, W. S. Hou, P. Chang {\it et al.} [Belle Collaboration],
 Nature 452 (2008) 332; and talk by P. Chang at this workshop.
\bibitem{Peskin} M. E. Peskin, Nature 452 (2008) 293.
\bibitem{Marco}  M.~Ciuchini, E.~Franco, G.~Martinelli, M.~Pierini and L.~Silvestrini,
  Phys.\ Lett.\  B{674} (2009) 197.
\bibitem{Li} H.~n.~Li and S.~Mishima,
  arXiv:0901.1272 [hep-ph].

\bibitem{Soni} See also A. Soni, A. K. Alok, A. Giri., R. Mohanta, S.
Nandi, arXiv:0807.1971 [hep-ph]; and talk by A. Soni at this
workshop.

\bibitem{partial-wave-unitarity}M.~S.~Chanowitz, M.~A.~Furman and I.~Hinchliffe,
Nucl. Phys. B{153} (1979) 402.
\bibitem{atlas-tdr}ATLAS Detector and Physics Performance Technical
Design Report. CERN/LHCC/99-14/15 (1999), section 18.2.
\bibitem{Chao} Talk by Y.~Chao at this workshop and the references therein.
\bibitem{Contino}   R.~Contino and G.~Servant, JHEP {0806} (2008) 026.
\bibitem{Ozcan} E.~Ozcan, S.~Sultansoy and G.~Unel, Eur. Phys. J. C57 (2008) 621;
and talk by E.~Ozcan at this workshop.
\bibitem{singlejet} B.~Holdom, JHEP {0708} (2007) 069; and talks by E.~Ozcan and B.~Holdom
at this workshop.

\bibitem{cakir} O.~Cakir, I.~Turk~Cakir, H.~Duran~Yildiz, R.~Mehdiyev,Eur.Phys.J. C
 56 (2008) 537; and talk by O. Cakir at this workshop.

\bibitem{Milstead}   M.~Fairbairn, A.~C.~Kraan, D.~A.~Milstead, T.~Sjostrand,
P.~Skands and T.~Sloan,   Phys.\ Rept.\  {438}, 1 (2007); and talk
by D. Milstead at this workshop.

\bibitem{belotsky}  K.~Belotsky, V.~A.~Khoze, A.~D.~Martin and M.~G.~Ryskin,
  Eur.\ Phys.\ J.\  C{36} (2004) 503; and talk by K. Belotsky at this workshop.
\bibitem{cuhadar} T.~Cuhadar-Donszelmann, M.~K.~Unel, V.~E.~Ozcan, S.~Sultansoy and
G.~Unel, JHEP {0810} (2008) 074; and talk by T.~Cuhadar at this
workshop.
\bibitem{cetin} E.~Arik, M.~Arik, S.A.~Cetin, T.~Conka, A.~Mailov and S.~Sultansoy,
Eur. Phys. J. C26 (2002) 9; and talk by S.~Cetin at this workshop.
\bibitem{haas} V.~M.~Abazov {\it et al.}  [D$\emptyset$ Collaboration],
Phys.\ Lett.\  B{663} (2008) 26; and talk by A.~Haas at this
workshop.

\bibitem{Vagnoni} Talk by V.~Vagnoni at this workshop.
\bibitem{Babarafb}
B.~Aubert {\it et al.}  [BaBar Collaboration], Phys.\ Rev.\  D 79
(2009) 031102.
\bibitem{Belleafb}
 J.~T.~Wei, P.~Chang {\it et al.}  [Belle Collaboration], arXiv:0904.0770 [hep-ex].
\bibitem{Serra} Talk by N. Serra at this workshop.
\bibitem{LHCbangular}
  U.~Egede, T.~Hurth, J.~Matias, M.~Ramon and W.~Reece,
  JHEP 0811 (2008) 032.


\bibitem{Richard} F.~Richard, arXiv:0807.1188 [hep-ph]; and talk at this workshop.
\bibitem{SultansoyFC} A.~K.~Ciftci, R.~Ciftci and S.~Sultansoy, Phys.\ Rev.\  D {72} (2005) 053006;
and talk by S. Sultansoy at this workshop.
\bibitem{SherFC} E.~De Pree, M.~Sher and I.~Turan, Phys.\ Rev.\  D {77} (2008) 093001;
and talk by M. Sher at this workshop.


\end{thebibliography}
\end{document}